\documentclass[
reprint,
superscriptaddress,
amsmath,amssymb,
aps,
prx,
longbibliography
]{revtex4-1}

\usepackage{graphicx,epsf}
\usepackage{float}
\usepackage{xcolor}
\usepackage{bm}

\begin{document}

\title{Magneto-conductance of topological junctions based on two-dimensional electron gases reveals Majorana phases}

%
\author{Lloren\c{c} Serra}
\affiliation{Institute of Interdisciplinary Physics and Complex Systems IFISC 
(CSIC-UIB), E-07122 Palma, Spain} 
\affiliation{Department of Physics, University of the Balearic Islands, 
E-07122 Palma, Spain}
\author{Kaveh Delfanazari}
\affiliation{Centre for Advanced Photonics and Electronics (CAPE), University of Cambridge, Cambridge CB3 0FA, UK}
\affiliation{Department of Physics, Cavendish Laboratory, University of Cambridge, Cambridge CB3 0HE, UK}

\begin{abstract}
We calculate the linear conductance of a two-dimensional electron gas (2DEG)-based junction between a normal semiconductor 
section and a hybrid 
semiconductor-superconductor section, under perpendicular magnetic field. 
We consider two important terms often neglected in the literature, the magneto-orbital and transverse Rashba spin-orbit. The strong orbital
effect due to the magnetic field yields topological phase transitions to nontrivial phases hosting Majorana modes in the hybrid section. The presence of a potential barrier
at the junction interface reveals the Majorana phases as quantized plateaus of high conductance, for low values of the chemical potential. 
In wide junctions (or large chemical potentials) the phase transitions occur at low magnetic fields but the magneto-conductance becomes anomalous and lacks clearly quantized plateaus.
\end{abstract}

\maketitle

\section{Introduction}

Majorana modes in nanostructures have been attracting strong interest since the experiments 
in semiconductor nanowires  gave initial evidence  on their real 
existence,\cite{Mourik,Das,Deng,Finck,Churchill} 
in agreement with earlier theoretical 
studies (see Refs. \onlinecite{Alicea12,Beenakker,Stanescu13,Aguado,Lutrev,Osc19} for reviews).
Additional experimental evidence has been also 
obtained more recently.\cite{Al16,Den16,Nic17,HaoZ,HaoZ2,deM18,Gri19,Bom19}
In semiconductor nanowires, Majorana states at the end points of the wire can be engineered
by combining three essential ingredients; spin-orbit interaction, magnetic field and superconductivity. While the 
spin-orbit coupling is intrinsic to the semiconductor, the other two ingredients are not;
superconductivity can be induced by proximity with a nearby bulk superconductor
and the magnetic field has to be tuned externally.
Although quantum wires are, ultimately,
idealizations of  1D systems, it soon became of interest to theorists the relevance of multi bands in quasi-1D (q1D) nanowires with transverse degrees of freedom; either 2D-q1D strips with a lateral width,\cite{Pot11,Lut11,Lim12b,Rai13,San14,Sed15} 
or 3D-q1D nanowires with a given shape of the transverse cross 
section.\cite{Groth14,Nij16,Man17,Stanescu19}  

The 2D-q1D geometry is of special interest, as it relates to semiconductor 2D electron gases (2DEG's) of widespread application
in semiconductor nanodevices. 
The induced supercondutivity in 2DEG's and planar Josephson junctions 
has been demonstrated in Refs.\ \onlinecite{Sha16,Kja16,Del17,Del18,Del19,For19,Ren19}.
In this planar geometry, a perpendicular magnetic field has a paramount influence on the motion of quasiparticles in the plane; the scenario of the well studied quantum Hall effect. 
In our context of hybrid semiconductor-superconductor systems,
the relevance of the magneto-orbital effect 
for the characterization of topological phases has been studied in different geometries;
cylinders,\cite{Lim13,Dmy18} 
faceted wires\cite{Nij16,Man17} and 2D strips or ribbons \cite{Osc15,Now18,Now18b}.
In 2D strips it is generally 
assumed that a perpendicular field is detrimental for the Majorana modes and a parallel field is 
more often considered where the magnetic effect is restricted to a Zeeman coupling with the quasiparticle spin. 
In Ref.\ \onlinecite{Osc15} 
it was shown that when tilting the field from the horizontal orientation towards the vertical orientation there are critical angles
beyond which observing the Majorana mode is no longer possible for weak SO coupling.
Remarkably, however, it was predicted that with stronger 
couplings (relative to the transverse confinement energy) there are parameter ranges where end Majorana modes may be present even in fully perpendicular fields. For a theoretical study, a disadvantage of the perpendicular field in 2D-q1D geometry is that the determination of the topological transitions is not known analytically but only numerically; approximate analytical limits 
require a few-band truncation.\cite{Now18} 
On the other hand, as mentioned above, the perpendicular field geometry is 
more convenient experimentally, since in this case the field has a maximal influence and thus 
lower fields are more effective for parameter tuning.

In this work we analyze the topological phase diagrams of 2D-q1D strips in presence of vertical field.
We show that the magneto-orbital effect leads to nontrivial Majorana phases for relatively low values of the 
field, depending on the chemical potential and intensity of the SO coupling. We then consider transport in 2D-q1D junctions between a normal semiconductor and a hybrid semiconductor-superconductor, with the purpose of identifying signatures of the nontrivial Majorana phases
of the hybrid strip in the linear conductance of the junction. We find that for low values of the chemical potential the junction conductance clearly reveals the topological phases as quantized conductance plateaus, robust against the presence of an interface potential barrier. On the contrary, the linear conductance of the trivial phases is severely quenched by an interface barrier. This clear 
difference in conductance between trivial and topological phases 
is restricted to relatively low chemical potentials, while it degrades for higher values.  We then consider wider junctions, where the 
SO becomes stronger relative to the transverse confinement energy, showing that the topological regions shrink toward zero field; i.e., 
the sequence of multiple transitions trivial-topological-trivial\dots takes place at 
lower fields than in narrow junctions.
On the whole, 
our results clarify the theoretical scenario for topological transitions
of hybrid 2DEG strips in vertical magnetic fields, specifically showing 
how the junction magneto-conductance reveals such topological phases.

\section{Model}

We use the model of a hybrid semiconducting-superconducting 2D-q1D wire
(as reviewed, e.g., in Ref.\ \onlinecite{Osc19}). Quasiparticle motion in the $xy$ plane is described with continuum
coordinates $(x,y)$, with $y$ restricted to 
$-L_y/2<y<L_y/2$, i.e., a strip of width $L_y$. Spin and electron-hole (isospin) degrees of  freedom are treated as discrete quantum variables with $\sigma_{xyz}$ and $\tau_{xyz}$ Pauli matrices, respectively. The Hamiltonian reads
\begin{eqnarray}
	\mathcal{H} &=& \left( \frac{p_{x}^{2}+p_y^2}{2m} -\mu \right)\tau_z 
+\frac{\alpha}{\hbar}\, \left(\, p_x \sigma_{y} - p_y \sigma_{x}\, \right)\tau_z	
	\nonumber\\ 
	&+& 
	\Delta_B\, \sigma_z + \Delta_0\, \tau_{x}\nonumber\\
	&+& \frac{\hbar^2}{2 m l^4_z}y^2 \tau_z 
	-\frac{\hbar}{m l^2_z}y\,p_x  
	- \frac{\alpha}{l^2_z}y \sigma_y
	 \; ,
\label{eq1}
\end{eqnarray}
where $\alpha$, $\Delta_B$ and $\Delta_0$ are the SO, Zeeman and 
pairing parameters, respectively. The Zeeman energy $\Delta_B$ is related to the field $B$ by
$\Delta_B=g^*\mu_B B/2$, where $g^*$ is the gyromagnetic factor.  
The last three terms of Eq.\ (\ref{eq1}), depending on the magnetic length
$l_z^{-2}=eB/\hbar c$,
are the orbital field terms. The chemical potential is represented by parameter $\mu$. 

We obtain below solutions of  Schr\"odinger's  stationary equation for a given energy $E$,
\begin{equation}
\left( \mathcal{H} - E \right) \Psi(xy\eta_\sigma\eta_\tau)= 0\;, 
\end{equation}
where $\eta_{\sigma,\tau}=1,2$ are the discrete spin and isospin variables, respectively.
We use the complex band structure approach, where the $x$ dependence of the the wave function is
expanded in a set of wave numbers $k$, including real (propagating) and complex (evanescent) 
modes. The $y$ dependence is described in a 1D grid of uniformly distributed points
and a general wave function is represented as 
\begin{equation}
\Psi(xy\eta_\sigma\eta_\tau)=
\sum_k{
C_k\, e^{ikx}\, \Phi_k{(y,\eta_\sigma\eta_\tau)}
},
\label{eq3}
\end{equation}  
where the $\Phi_k$'s are determined from the solution of the 1D $k$-dependent eigenvalue problem
(see below) and the $C_k$'s are the set of complex amplitudes representing a given 
state.\cite{Osc19}

The $\Phi_k$'s in Eq.\ (\ref{eq3}) are the transverse states of an 
infinite homogenous strip. They characterize the strip band structure $\varepsilon(k)$, for real values of $k$, from the eigenvalue equation 
\begin{equation}
h_k \Phi_k(y\eta_\sigma\eta_\tau)=\varepsilon_k\Phi_k(y\eta_\sigma\eta_\tau)\; ,
\label{eq4n}
\end{equation}
where the 1D Hamiltonian $h_k$ is obtained by the replacement $p_x\to\hbar k$ in the general 
Hamiltonian ${\cal H}$ of Eq.\ (\ref{eq1}). The strip topological transitions are characterized by the 
$k=0$ gap closings,\cite{Lutchyn,Ore10} i.e., by the condition $\varepsilon_0=0$. 
Specifically, we will analyze below the $B$-$\mu$ phase diagram
of the topological strip determining the $\varepsilon_0=0$ curves in the diagram. 
Besides, it is also of interest to determine the regions where the global gap, i.e., the eigenvalue $\varepsilon_k$ for some $k$ not necessarily zero, vanishes.\cite{Kli12} 
This global gap vanishing condition $\varepsilon_k=0$ defines sizeable portions of the $B$-$\mu$ plane, the gapless 
regions, in addition to the mentioned $\varepsilon_0=0$ curves of zero measure in the plane. In the gapless
regions there are propagating states at zero energy and, therefore, any localized zero mode like the Majorana mode will decay into those extended states. 

\begin{figure}[t]
\begin{center}
\includegraphics[width=0.375\textwidth,trim=3cm 3.3cm 6cm 0.75cm, clip]{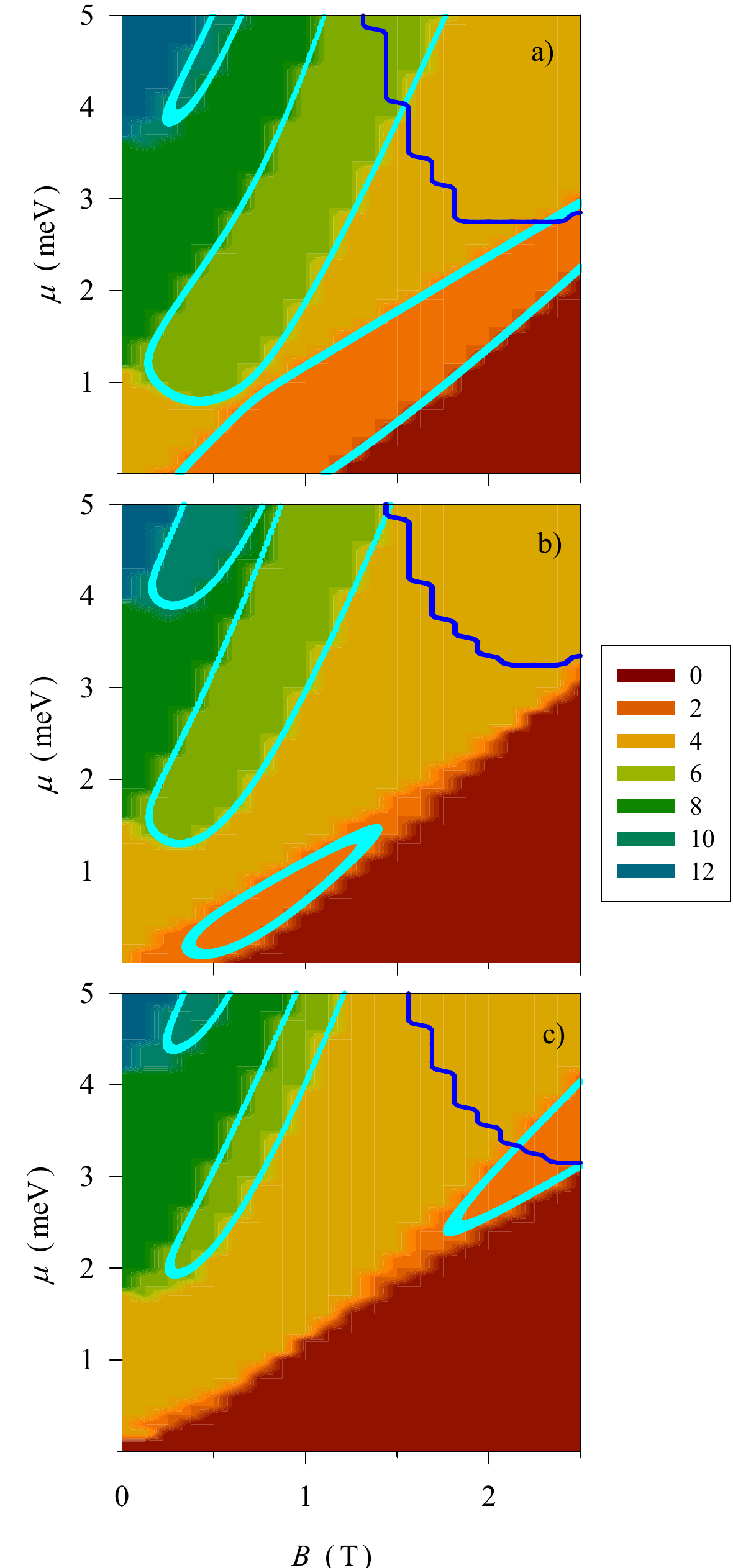}
\end{center}
\caption{Phase diagrams of a hybrid  strip with $\Delta_0=0.3\;{\rm meV}$, $L_y=150\; {\rm nm}$ and for different values of the SO coupling: $\alpha=60\; {\rm meV}\,{\rm nm}$ (a), 
$45\; {\rm meV}\,{\rm nm}$ (b) and 
$30\; {\rm meV}\,{\rm nm}$ (c).
The cyan lines encircle the topological regions with a Majorana mode in the 
$\mu$-$B$ plane, obtained as gap closings for $k=0$.
The dark blue line shows the boundary for activation of a propagating mode, separating the region of 
any-$k$ gap closing  in the upper right corner.  
The background color indicate the number of electron/hole propagating modes  in the corresponding normal 
strip having $\Delta_0=0$ and the same rest of parameters ($N_{eh}\equiv 2N$).
Additional parameters: $m=0.033\, m_e$, $g^*=15$. 
}
\label{Fig1}
\end{figure}
   
In practice, we have determined the eigenvalues and eigenstates of Eq.\ (\ref{eq4n}) numerically, 
discretizing the $y$ coordinate in a uniform grid and using sparse matrix
diagonalization techniques.\cite{arpack} 
This method is very efficient computationally and allows
well converged results with the use of large enough grids ($\gtrapprox 100$ points). 
In the junction case the Hamiltonian parameters are no longer constant 
since
we assume a vanishing pairing $\Delta_0=0$ for $x<0$; corresponding to the
normal semiconductor. The scattering problem is then solved as in Ref.\
\onlinecite{Osca2017}, determining the transmission
and reflection probabilities corresponding to a given incident mode. Adding the contributions from all possible incident modes we determine the linear conductance $G$ as\cite{Blo85}
\begin{equation}
G= \frac{e^2}{h}\left(\,N-R+R_A\right)\; ,
\label{eq4}
\end{equation}
where $N$ is the number of incident electron modes, $R$ is the normal reflection probability and $R_A$ is the Andreev reflection probability. Equation (\ref{eq4}) contains the well-known result that
while normal reflections decrease the conductance, 
Andreev reflection processes, whereby incident electron quasiparticles are reflected as holes, enhance the 
conductance.

\section{Results}

\subsection{Phase diagrams}

Figure \ref{Fig1} shows the $B$-$\mu$ phase diagrams of a hybrid strip of $L_y=150$ nm 
for different values of the SO coupling $\alpha$.
Since the value of $\alpha$ is sample dependent  and it can be actually tuned 
with electric fields we have chosen some representative values within the typical 
range of InAs 2DEG's with the purpose of investigating the tendency for increasing or 
decreasing $\alpha$.\cite{Tak17}  
The topological
regions with one  Majorana mode are indicated by the cyan lines in Fig.\ \ref{Fig1}.
Considering, for instance, an evolution of parameters with constant $\mu$ and increasing $B$ from the trivial $B=0$ phase, there is a phase transition to a one-Majorana 
phase whenever a cyan line is first crossed. That is, the cyan lines are 
enclosing one-Majorana phases in a background of trivial phase. A similar topology of the $B$-$\mu$ plane was already obtained in Ref.\ 
\onlinecite{Die12} but neglecting the terms in 
$\alpha p_y$ and all the orbital $l_z$ terms.
We find that the shape of the transition borders 
is strongly affected by these terms, that actually dominate for large enough fields.
Remarkably, we also find a {\em Majorana island} surrounded by trivial phase in Fig.\ \ref{Fig1}b. 

There is a clear correspondence in Fig.\ \ref{Fig1} between one-Majorana regions
and regions with an odd number $N$ of electron propagating modes in the normal strip 
(the strip with $\Delta_0=0$ and the same rest of parameters). The latter are shown by the colorscale background in Fig.\ \ref{Fig1}, only minor deviations at very small fields can be observed between both types of regions. This relation is not obvious a priori and it was already pointed out
in Ref.\ \onlinecite{Die12} as a correspondence between the topological number $|Q|$ of the hybrid system
and the number of propagating modes $N$ of the normal one. We find that  such correspondence is thus preserved by the magneto-orbital terms.

The boundary for the activation of 
propagating modes in the hybrid system is indicated by the blue lines in Fig.\ \ref{Fig1}; the gapless
phase corresponding to the upper right corner of each panel. In this gapless phase the Majorana mode 
looses its protection as it decays into propagating modes with the same zero energy. As mentioned above, 
the gapless phase is a sizeable region where propagating modes exist for a specific finite (nonvanishing) $k$
and it should be distinguished from the cyan lines in Fig.\ \ref{Fig1} representing gap closing only for $k=0$. 

\subsection{The magneto-conductance plateaus}

\begin{figure}[t]
\begin{center}
\includegraphics[width=0.3\textwidth,trim=1cm 9cm 4.5cm 1.cm, clip]{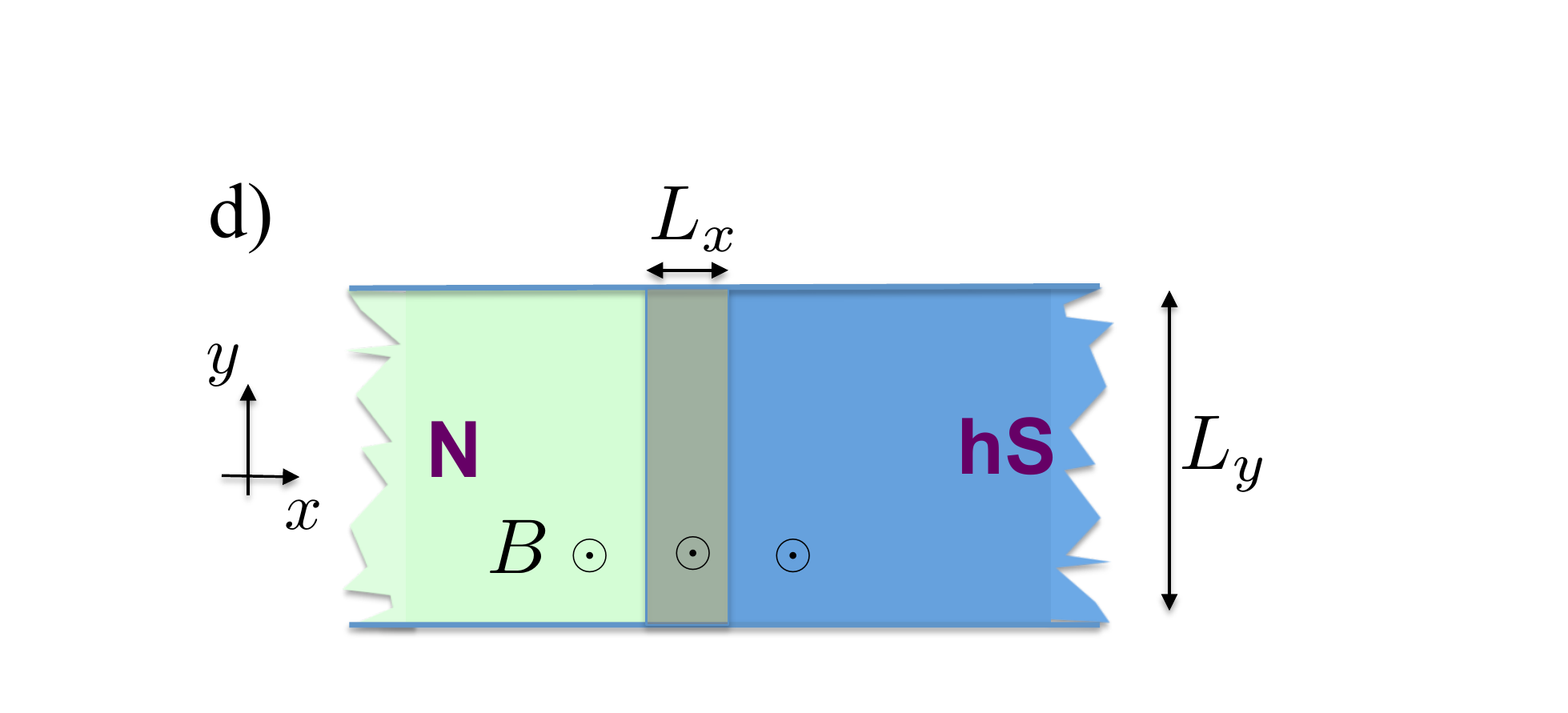}
\includegraphics[width=0.4\textwidth,trim=2cm 7cm 2.5cm 1.cm, clip]{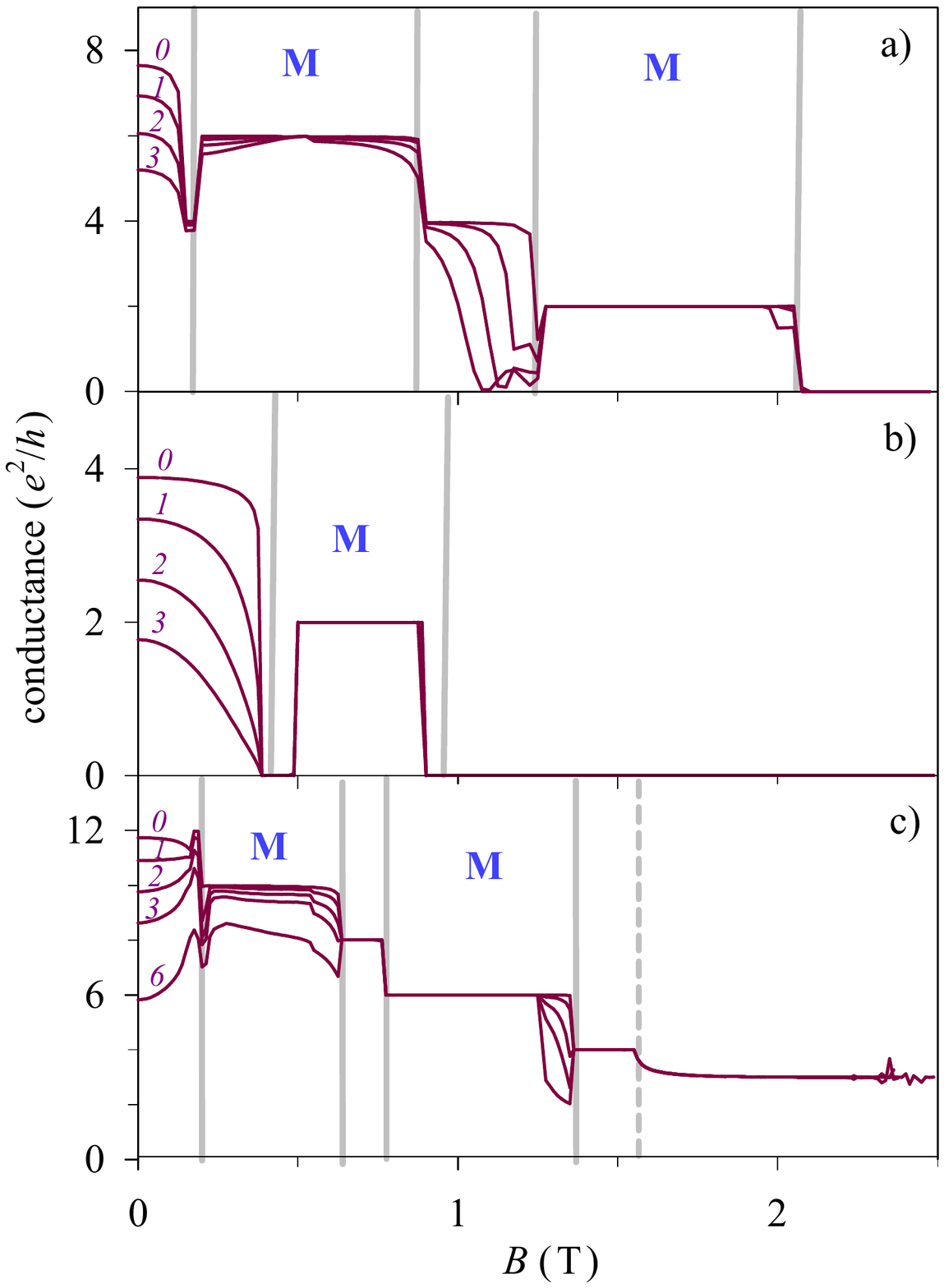}
\vspace{-3mm}
\end{center}
\caption{ d) (upper)
Sketch of a junction with a potential barrier at the interface.
a,b,c) (lower)
Junction conductance as a function of the field for $L_y=150\; {\rm nm}$.
Panel a) is for $\alpha=60\,{\rm meV}\,{\rm nm}$ and $\mu=1.5\,{\rm meV}$; panel b) is for $\alpha=45\,{\rm meV}\,{\rm nm}$ and $\mu=0.5\,{\rm meV}$,
and panel c) is for $\alpha=45\,{\rm meV}\,{\rm nm}$ and 
$\mu=4.5\,{\rm meV}$.
The different curves  
are for increasing barrier heights at the interface, 
with the corresponding value of $V_0$ in meV indicated near each curve. 
Solid vertical bars show the phase transition
boundaries from Fig.\ \ref{Fig1}, while the dashed bar in panel c) signals the activation 
threshold for propagating modes (blue curves in Fig.\ \ref{Fig1}).
Majorana phase regions are indicated with an $M$.
Parameters: $L_x=7\; {\rm nm}$, $g=15$, $m=0.033 m_e$.
}
\vspace{-5mm}
\label{Fig2}
\end{figure}

We turn now to calculating the magneto-conductance $G(B)$ when a normal lead is attached
to a hybrid semi infinite strip, forming a 2D-q1D junction (see sketch in Fig.\ \ref{Fig2}d). 
We will focus on the the small bias (linear) regime obtaining the conductance from Eq.\ (\ref{eq4})
and including a potential barrier $V_0$ of length $L_x$ near the junction
edge in order to
tune the effective coupling between the N and hS parts. 
We choose a fixed value of
$\mu$ and then sweep $B$ in order to probe the different system phases displayed in Fig.\ \ref{Fig1}. The results of Fig.\ \ref{Fig2} show some representative cases; panel a) 
corresponds to $\mu=1.5\; {\rm meV}$ in Fig.\ \ref{Fig1}a, panel b) to $\mu=0.5\; {\rm meV}$ in Fig.\ \ref{Fig1}b and
panel c) to $\mu=4.5\;{\rm meV}$ in Fig.\ \ref{Fig1}b. The different conductance traces 
in each panel
correspond to increasing barrier heights $V_0$ at the interface; higher $V_0$ corresponding to the lower value of the conductance.
The vertical grey bars in Fig.\ \ref{Fig2} are the topological region boundaries from Fig.\ \ref{Fig1}. 

It is remarkable that the interface barrier $V_0$ strongly quenches the conductance of the trivial regions in Figs.\ 
\ref{Fig2}a and \ref{Fig2}b, while it leaves 
almost unaffected the conductance of the topological phases. This behavior can be explained by the strong 
sensitivity of the Andreev reflection to the barrier in the trivial phases. On the other hand, the presence of a Majorana mode in the topological phases leads to more robust quantized conductance plateaus. 
The topological robustness against an interface barrier provides an 
interesting experimental way to discern the 
presence of a Majorana mode. This appealing scenario, however, is valid only for relatively low chemical potentials. As shown in Fig.\ \ref{Fig2}c, with larger  chemical potentials the barrier sensitivity is no longer resolving well the trivial and topological phases; rather, it yields a different behavior at low and high fields, 
irrespectively of the topological phase in each case.
At low fields the conductance is always quenched while at high fields it always remains quantized. 
Besides, with larger $\mu$ there are propagating modes 
induced by the magnetic field beyond a certain value, indicated by the dashed line in 
Fig.\ \ref{Fig2}c. This opening of transmision channels causes a
counterintuitive effect; a conductance decrease due to the decrease of 
Andreev reflection $R_A$.

\begin{figure}[t]
\begin{center}
\includegraphics[width=0.475\textwidth,trim=1cm 18.cm 1.5cm 1.cm, clip]{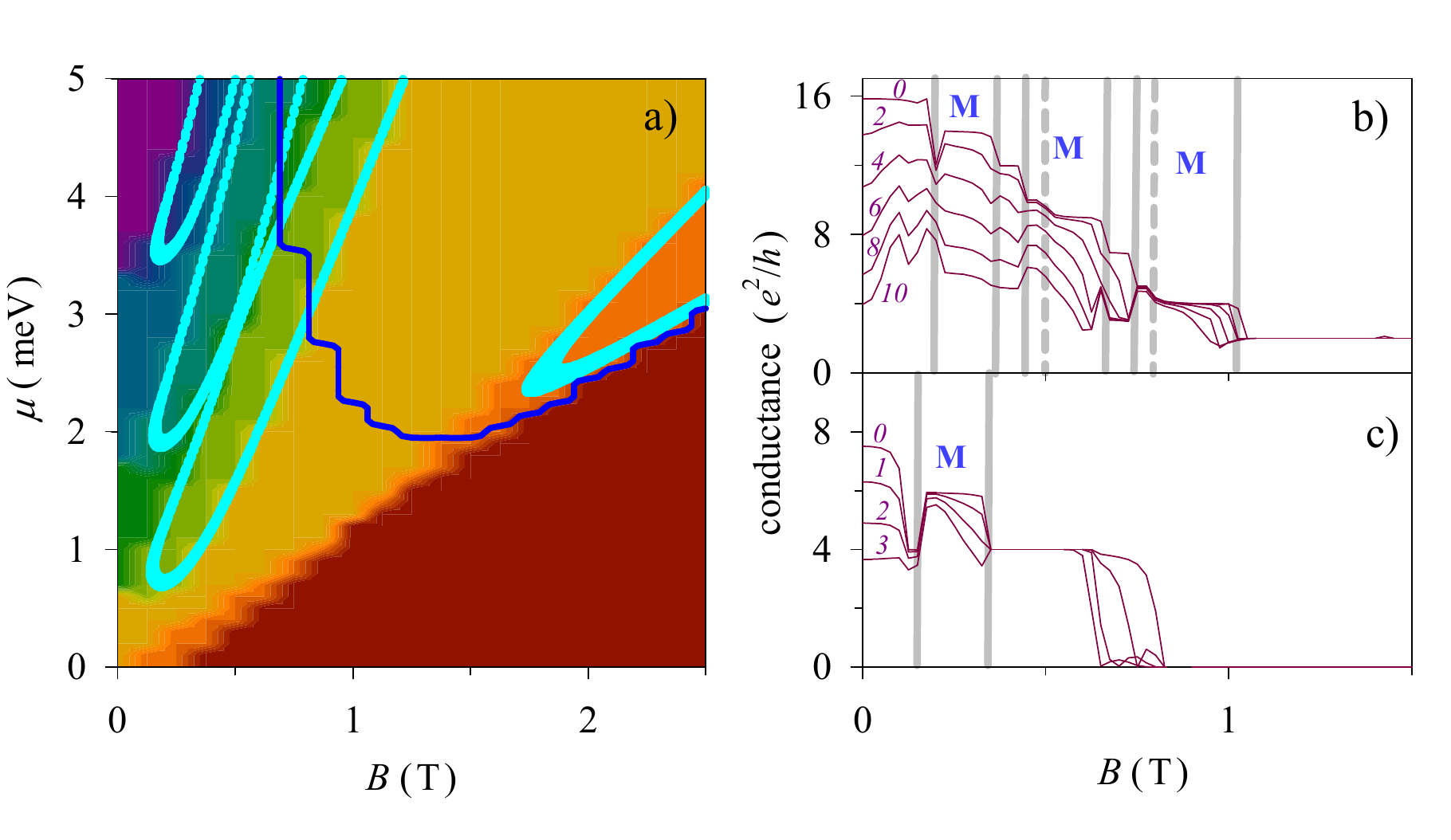}
\vspace{-3mm}
\end{center}
\caption{a) Topological phase diagram similar to Fig.\ \ref{Fig1}c but for a wider junction of 
$L_y=225$ nm. 
The right panels show the junction conductance for $\mu=4\; {\rm meV}$  (b) and 
$\mu=1\;{\rm meV}$ (c). As in Fig.\ \ref{Fig2} the results for increasing potential barriers are displayed from 
upper to lower curves and the value of $V_0$ in meV is given.
The Majorana phases and the activation thresholds for propagating modes 
are also indicated as in Fig.\ \ref{Fig2}.
Parameters: $\alpha=30\, {\rm meV}\,{\rm nm}$,
$L_x=7\, {\rm nm}$.
}
\vspace{-5mm}
\label{Fig3}
\end{figure}

\subsection{Dependence on $L_y$}

We consider next the influence of varying $L_y$ on the above results. 
In wider junctions the effective energy scale set by the transverse confinement is reduced and, therefore, 
a fixed $\mu$ and $\alpha$ values will evolve from weak to strong regime by simply increasing $L_y$. 
The phase diagram
for $L_y=225$ nm is shown in Fig.\ \ref{Fig3}.
We notice that, within the explored window, the tendency with increasing $L_y$ is to shrink a sequence of topological regions towards lower fields, from two regions with $B<1.5\; {\rm T}$ in Fig.\ \ref{Fig1}c to three regions in Fig.\ \ref{Fig3}a.
The gapless part, shown by the blue line, is enlarged with respect to Fig\ \ref{Fig1}, thus leaving the protected Majorana phases only for rather small fields. Therefore, there is a severe compromise when
increasing $L_y$ between the sequence of low field topological transitions and the gap closing for propagating modes.

Figures \ref{Fig3}b and c show the conductance traces for two selected values of the chemical potential of Fig.\ \ref{Fig3}a. In these cases, 
the effect of the interface barrier is not clearly identifying the Majorana phases. This is similar to the 
result in Fig.\ \ref{Fig2}c, indicating that with larger $L_y$'s the 
identification of robust Majorana plateaus as in Fig.\ \ref{Fig2}ab
is not possible by simply increasing $V_0$. This can be understood noting that the larger $L_y$ implies a lower confinement energy and, thus, the requirement of low values of chemical potential already obtained in Fig.\ \ref{Fig2} becomes even more restrictive for Fig.\ \ref{Fig3}.  
We want to stress, however, that even in cases when the interface barrier is not yielding perfectly quantized plateaus in the topological regions, 
the conductance still manifests a non monotonous $B$ dependence, usually with minima before the onset of a topological region, in Fig.\ \ref{Fig3}bc. This {\em anomalous} magneto-conductance at low fields, as compared to a smooth decrease due to the 
magnetic depopulation of propagating bands, still originates
in the topological transitions of the hybrid system.

\begin{figure}[t]
\begin{center}
\includegraphics[width=0.475\textwidth,trim=1cm 18.cm 1.5cm 1.cm, clip]{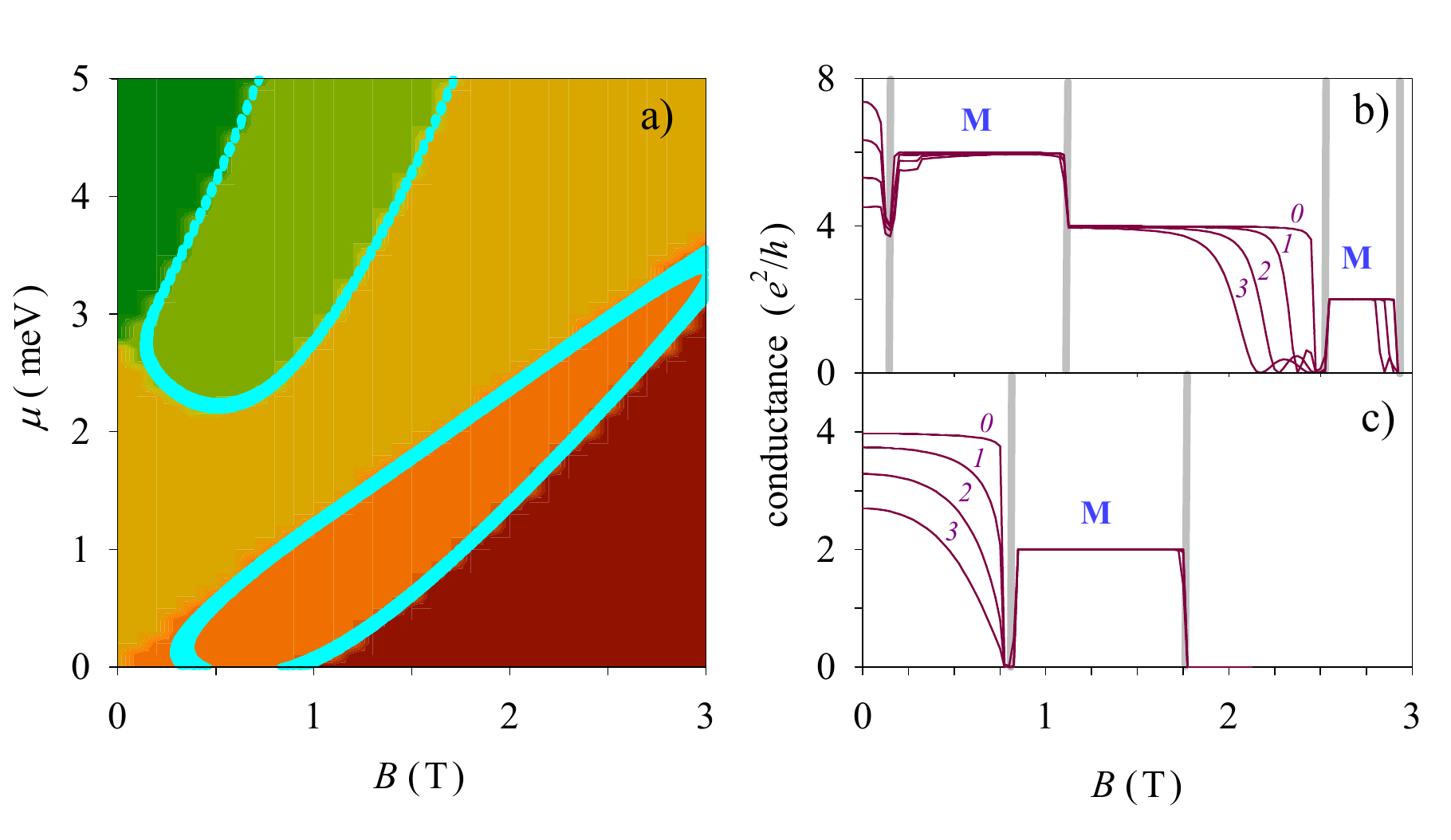}
\vspace{-3mm}
\end{center}
\caption{ Same as Fig.\ \ref{Fig3} for a narrower junction of width $L_y= 112$ nm. Panels b) and c) 
show the magneto-conductance curves for $\mu=3\, {\rm meV}$ and 1 meV, 
respectively. Parameters: $\alpha=60\; {\rm meV}\,{\rm nm}$, $L_x=7\,{\rm nm}$.
}
\vspace{-5mm}
\label{Fig4}
\end{figure}

With a smaller width, $L_y=112\; {\rm nm}$ in Fig.\ \ref{Fig4},
the magneto-conductance scenario of topological quantized plateaus
and quenched trivial regions is reinforced. In this case, the hybrid
part is always gapped and 
the Majorana regions are more separated, for the inspected 
window of parameters in Fig.\ \ref{Fig4}a. 
Even the relatively larger chemical potential $\mu=3\;{\rm meV}$
is showing a quenched conductance around $B\approx 2\;  {\rm T}$ in Fig.\
\ref{Fig4}b followed by a sharp increase when entering the topological 
phase around $B\approx 2.6\; {\rm T}$.

\section{Conclusions}

We have calculated the phase diagrams of hybrid semiconductor-superconductor 
2DEG strips in vertical magnetic field. The $\mu$-$B$ plane contains Majorana regions
that depend on the values of SO coupling ($\alpha$) and transverse width ($L_y$). For increasing $L_y$'s or increasing $\alpha$'s the topological regions are squeezed towards lower fields. We have then investigated the  magneto-conductance of 
the junction between a normal and a hybrid strip, in presence of a potential barrier at the interface. For low values of the chemical potential 
we find a 
magneto-conductance 
scenario of robust quantized topological plateaus and quenched trivial regions,
with respect to an interface potential barrier. More in general, for higher chemical potentials or transverse widths, we find that the sequence of topological regions with increasing field causes a nonmonotonous magneto-conductance,
with anomalous steps and minima usually before the onset of topological regions, 
in presence of an interface barrier. 

Our results suggest a direct validation of the topological phases in 2D 
junctions by means of magneto-conductance measurements. The strong 
orbital magnetic effect in perpendicular field can be used advantageously 
to engineer topological transitions at relatively low fields in
hybrid semiconductor-superconductor 2DEG strips. We calculated a
lowest critical field of $B_c\approx 0.2\,{\rm T}$ with the parameters of Fig.\ 
\ref{Fig3}, but even lower $B_c$'s would be achieved with 
smaller pairings $\Delta_0<0.3\; {\rm meV}$ and wider junctions
$L_y>225\,{\rm nm}$. 

\begin{acknowledgements}
We acknowledge support from MINECO (Spain), grant
MAT2017-82639, and from EPSRC (UK).
\end{acknowledgements}

\appendix

\begin{figure}[t]
\begin{center}
\includegraphics[width=0.45\textwidth,trim=1cm 11.cm 4cm 0.cm, clip]{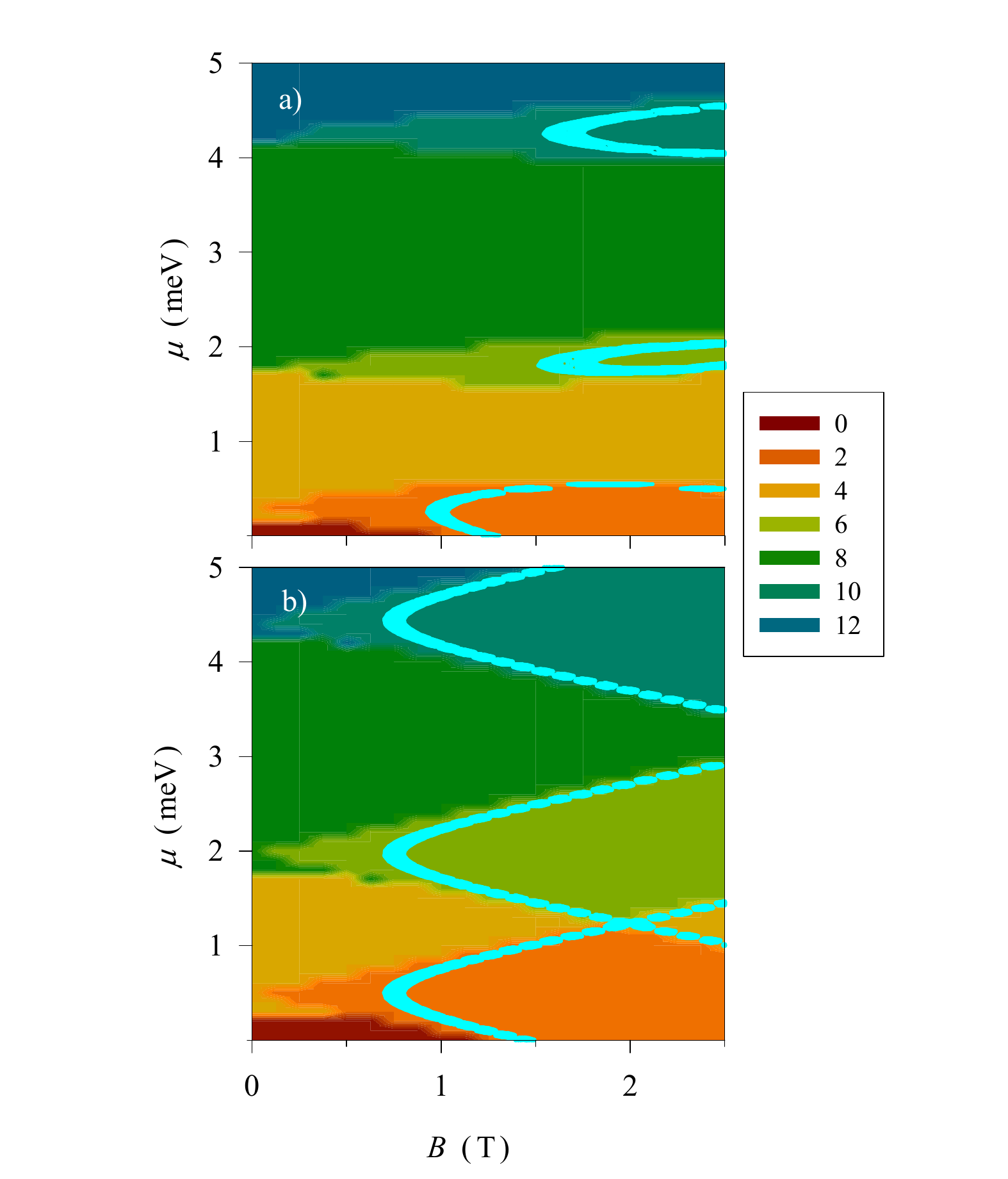}
\vspace{-3mm}
\end{center}
\caption{
Phase diagrams similar to Fig.\ \ref{Fig1} but neglecting the magneto-orbital terms in the Hamiltonian, i.e., neglecting the last three contributions of Eq.\ (\ref{eq1}). Panel a) is for the full SO interaction $(p_x\sigma_y-p_y\sigma_x)$ while panel b) is for the  case of a partial SO containing only the longitudinal $p_x\sigma_y$ contribution. Parameters:
$L_y=150\; {\rm nm}$, $\alpha=30\; {\rm meV}{\rm nm}$.
}
\vspace{-5mm}
\label{Fig5}
\end{figure}

\section{Phase diagrams without orbital effects}

Since the magneto-orbital terms are often neglected in the literature, it is worth confronting our above results for the full Hamiltonian, Eq.\ (\ref{eq1}),  with the 
case of total absence of orbital contributions; that is, neglecting the last three terms depending 
on the magnetic length $l_z$ in Eq.\ (\ref{eq1}). Another frequent simplifying assumption 
in the literature
is neglecting 
the transverse Rashba SO term $\propto\alpha p_y\sigma_x$ in Eq.\ (\ref{eq1}). This transverse SO term is the source in multiband wires for the coupling between Majorana modes of different bands,
leading to an effective repulsion between different Majorana regions
of the phase diagram.

Figure \ref{Fig5} confirms that in absence of orbital effects the Majorana phases, highlighted by the cyan lines,  appear at larger fields. Comparing Fig. \ref{Fig5}a to the corresponding Fig.\ \ref{Fig1}c  with the same 
$\alpha$ and $L_y$, we notice the high relevance of the orbital effect on the phase diagram. While the absolute lowest critical field in 
Fig.\ \ref{Fig1}c is $B_c\approx 0.2\; {\rm T}$ achieved
for $\mu\approx 2$ and $4.5\; {\rm meV}$, in  
Fig.\ \ref{Fig5}a it is $B_c\approx 1\; {\rm T}$ for $\mu\approx 0.2\; {\rm meV}$.
Thus, orbital field effects can not be neglected in a quantitative study of the critical fields of 
hybrid strips.
It is also worth stressing that the whole window of parameters represented in Fig.\ \ref{Fig5}a corresponds to a gapped spectrum (excluding, of course, the $k=0$ gap-closing lines of the phase transition boundaries); while in Fig.\ \ref{Fig1}c there is a sizeable gapless region highlighted by  a blue line. The orbital terms thus favour the gap closing of the spectrum in larger regions, lowering the 
parameter ranges where Majorana protection is ensured.  
We have also checked that  similar modifications are obtained neglecting orbital terms in the other two panels of Fig.\ \ref{Fig1} for the larger values of $\alpha$.

Neglecting the Rashba transverse term $\alpha p_y\sigma_x$, in addition to the orbital terms, leads to the phase diagram of Fig.\ \ref{Fig5}b. In this case, the Majorana region boundaries are analytical\cite{Lutchyn,Ore10}
\begin{eqnarray}
\!\!\!
\Delta_B &=& \sqrt{\Delta_0^2+\left(\mu-\frac{\hbar^2\pi^2 n^2}{2mL_y^2}\right)^2}\; , \;\;
n=1,2,\dots\; ,
\end{eqnarray}
the condition for the absolute lowest critical field being  $\Delta_B=\Delta_0$ for values of the  chemical potential equal to the successive transverse eigen energies $\mu=\hbar^2\pi^2n^2/(2 m L_y^2)$
with $n=1,2,\dots$. 
Since we assumed $\Delta_0=0.3\;{\rm meV}$,
this corresponds
in Fig.\ \ref{Fig5}b to $B_c\approx 0.7\; {\rm T}$ for $\mu\approx (0.5, 2, 4.5, \dots)\; {\rm meV}$. 

\bibliography{NSpap} 

\end{document}